4

# Register-Like Storage Block Used as Cluster Buffers, Histograms, and Hough Transform Accumulators for HEP Trigger Systems

Jinyuan Wu

*Abstract*— In high energy physics experiment trigger systems, block memories are utilized for various purposes, especially in binned searching algorithms. In these algorithms, the storages are demanded to perform like a large set of registers. The writing and reading operation must be performed in single clock cycle and once an event is processed, the memory must be globally reset. These demands can be fulfilled with registers but the cost of using registers for large memory is unaffordable. Another common requirement is the boundary coverage feature during reading process. When a memory bin is addressed, the stored contents in the addressed bin and its neighboring bin must be output simultaneously. In this paper, a register-like block storage design scheme is described, which allows updating memory locations in single clock cycle, reading two adjacent bins, and effectively refreshing entire memory within a single clock. The implementation and test results are presented.

## I. Introduction

IN modern high energy physics experiments trigger systems, data are usually reorganized in binned memories, or cluster buffers. Consider a multilayer detector as illustrated in Fig. 1, hit data from each detector layer arrive in random order and are to be stored in a binned memory for future reading out.

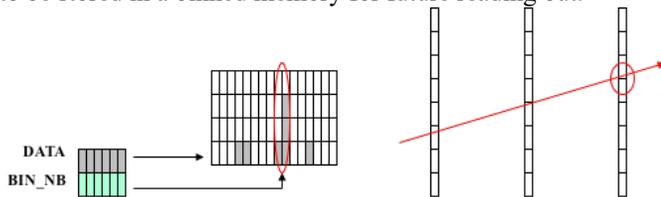

Fig. 1. An application of block memory bins used for in high energy physics experiment trigger system.

In this example, charged particle tracks pass through several detector layers. The data from the particle hits are to be clustered together so that tracks can be reconstructed for further trigger algorithm. We consider a sub-process of using a track segment candidate based on the hits from the first two detector layers to find the hit data on the third layer.

In the first step of this sub-process, detector raw data are stored in the memory bins. A bin in this example represents a range of coordinate on the third detector surface and in many cases, a bin may include several detector elements such as silicon strips, scintillating fibers or straw tubes. Therefore, in an event, it is possible to have multiple hits within a bin. Once all hits in an event are written into the memory, they are to be read out according to the location being pointed by the track segment.

A crucial requirement is the boundary coverage during the readout process. The track segment points to a location on the surface of the detector and an integer is derived from the location coordinate as the bin number (BIN_NB) which is used to address the memory. It is very common that the track segment points to a location near a boundary of a bin and the actual hit data may be store either in the pointed bin or in the neighboring bin. It is usually necessary to bring hit data from both bins out for finer track fitting processes.

The detector data are fed into the trigger system and in the processing stages of the firmware, the data are usually fetched one hit per clock cycle. The memory block bins are to be updated as the data fetched in every clock cycle. While writing a data into a memory within one clock is not difficult, it is a challenge to update the memory location within a single clock cycle. To update a memory bin, the contents of the bin must first be read out and the new hit data is concatenated into the original data word to form a new data word which is then written back into the memory bin. The updating process takes several clock cycles to complete which requires a dual port memory with a reading port and a writing port and a suitably designed pipeline so that the data can be processed one hit per clock cycle. Note that once a hit to be filled into a bin is fed into the pipeline, another hit to be filled into the same bin can come as early as the next cycle. In this case, the first data has not been written back into the memory bin before the reading cycle of the second update process. This is similar as the read-after-write (RAW) hazard in contemporary microprocessor design. To solve this hazard, a data forwarding scheme [1-3] is utilized and we will discuss the detail in later sections.

The trigger firmware processes data in event-based fashion, and an "event" in collider experiments is usually a beam crossing. The process takes three phases: (1) storing data or "booking", (2) reading data and (3) refreshing the memory. The input hit data are first fed in one word per clock cycle and are stored into the memory bins. After filling up the memory with the data from an event, the trigger algorithm will search the data based on the index of the bins and read them out. Note that the searching process may not address all bins containing the hit data and it may also address empty bins. After reading process, a single clock refreshing command is issued by the users and all memory bins will be effectively cleared to prepare for the next event. It is well known that regular block memories do not support global reset. To fulfill this requirement, an event ID tagging scheme is used.

The Hough transform is a popular track segment seeding scheme [4-12], in which an array of accumulators is utilized to book a 2D "histogram" for each coming detector hit. Using logic elements in FPGA to implement the Hough transform accumulator would consume large amount of silicon resources. Although the Hough transform accumulators do not look like clustering buffers described above, intrinsically they share many common behavior features. Using register-like storage block to implement Hough transform accumulators will help to reduce silicon area usage significantly.

The single clock operation and global refreshing schemes developed in our previous work [1-3] are combined into a unified scheme in this work. In this paper, the full structure of the register-like storage block is first discussed in Section II, followed by implementation and test results Section III. In Section IV, we will describe several applications of the register-like storage block including cluster buffers, histograms and an actual implement of a Hough transform accumulator array for track segment seeding applications.

## II. STRUCTURE OF THE REGISTER-LIKE STORAGE BLOCK

To fulfill the requirements of single clock updating, reading and refreshing operations, the functional block is organized as a set of pipelines. The block diagram of register-like storage block is shown in Fig. 2. It can be unfolded based on its operating logic for clarity as shown in Fig. 3.

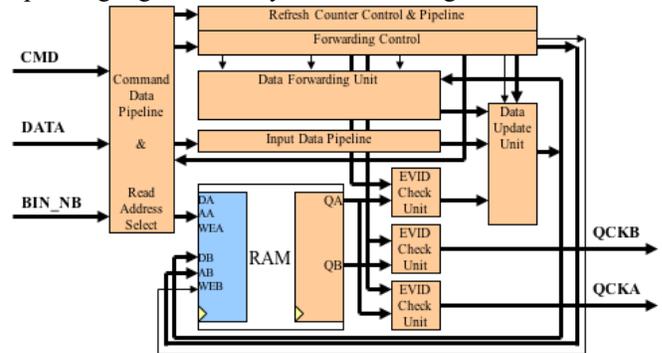

Fig. 2. Block diagram of the register-like storage block

During the process of data booking, data to be stored (DATA), the operating command (CMD) and the bin index (BIN_NB) arrive to the input ports of the pipeline at the same clock cycle. The bin index is selected to feed into port AA to read out the contents stored in the memory bin. After two clock cycles, the contents appear at port QA to be checked in the EVID Check Unit. Several bits in the memory words are assigned as event ID field, if the current location was written during the booking process in the current event, this field will store the current event ID. Otherwise, it will store the event ID from old events. The event ID is checked with the current event ID which is maintained in Refresh Counter Control & Pipeline block. If the stored contents are from the current event, it will be sent to the Data Update Unit for further process. If not, the stored contents will be ignored and all output bits can be optionally set to zeros.

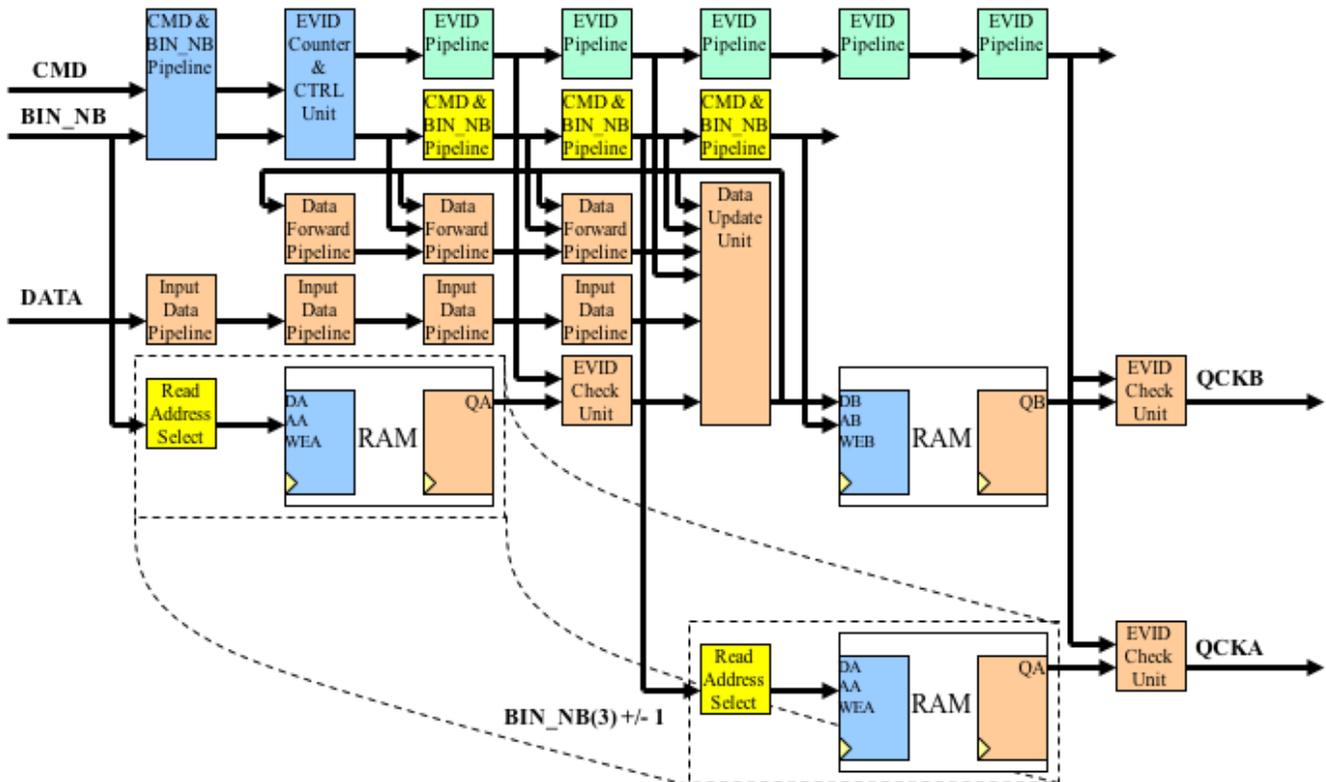

Fig. 3. Logic structure of the register-like storage block

In the Data Update Unit, the new input data (after appropriate pipeline delay) and the old contents in the memory bin are used to build a new data word. The updating algorithm can be chosen by the users depending the application. For example, the old contents can be moved upward to higher bits and the new data will be inserted to the lower bits of the data word, which will allow the memory bin to store multiple hits. For another popular application, histogram booking, the new input data is always 1 (so it is unnecessary to implement the Input Data Pipeline), the old content is simply added by 1.

The output of the Data Update Unit is sent to the second memory port DB to write back to the memory bin. At the same time, the result is also sent to the Data Forwarding Unit. The Forward Control block checks input data bin number to determine whether the Data Update Unit should use the data read out from the memory or there is a newer data in the Data Forwarding Unit. The Data Forwarding Unit always contains the most current values of the memory bins that are just updated.

After data booking process, the data reading out process uses the same pipeline and the read address is delayed and sent to port AB and the contents appear at QB, which are checked by another EVID Check Unit.

The reason of delaying the read out process is to wait for the memory bins fully updated. In this arrangement, the last writing command can be immediately followed by the first reading command without losing a clock cycle, which is crucial for high luminosity trigger systems.

Note that during the reading process, output contents from memory port QA is not used for Data Update Unit as in the booking process. Therefore, the A port of the memory is free to be used to read out contents in any bin. While port BA is addressed with BIN_NB (after appropriately delayed in the pipeline), we generate a new integer BIN_NB + 1 or -1 to address AA. This way, the outputs from QA and QB contain raw data from two neighboring bins allowing suitable boundary coverage.

After the reading process, a single clock cycle REFRESH command is issued which overwrite a rotationally selected memory bin (after several cycles of pipeline delay) and changes the EVID for the new event (or beam crossing in some applications). The booking command for the next event can be issued immediately following the refresh cycle. The book, read and refresh processes for an event can be connected end-to-end together without any missing clock cycles.

It should be point out that during the booking and the reading processes, the logic position of the memory port A in the pipeline is different.

During the booking process, the port A is used to read out the old contents in a bin stored in the memory. Therefore, it is addressed with the BIN_NB immediately at the early stage in the pipeline. For reading process, however, the port A is used to read out data in the neighboring bin. In this case, it is addressed with a delayed version of BIN_NB +1 or -1 and its logic position in the pipeline becomes the same as the port B.

### III. IMPLEMENTATION AND TEST RESULTS

The register-like storage block is designed, implemented, and tested in an Altera Cyclone V FPGA device (5CEBA4F23C7N). The actual top design interface block is shown in Fig. 4. The block uses 466 logic elements (ALM) which is about 3% of the target device.

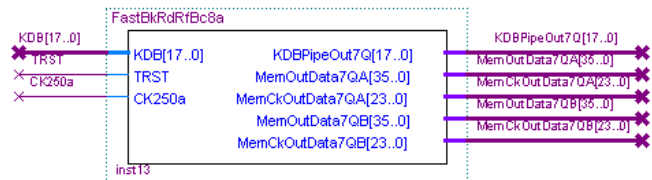

Fig. 4. The actual top design interface block of the register-like storage block

In this demo design, the input hit data is chosen to be 8-bit for simplicity. The block RAM are divided into 256 memory bins with 36 bits each that can store up to 3 hits and a 12-bit event ID. The block is driven by a 250 MHz clock which is generated with a phase lock loop (PLL) inside the FPGA device. In every clock cycle, an 18-bit word (KDB[17..0]) containing a 2-bit command, an 8-bit data and an 8-bit bin number is fed into the block. The 2-bit command word instructs the block to operate one of the four processes: NOOP, Book, Read and Refresh.

In book operation, the input data is filled into the memory indexed by the bin number while in read operation, the data word is primarily ignored and only the bin number is used. In our design, bit 0 of the data word is used as the "half-bin" bit of the bin number which is used to indicate if the actual index is closer to higher or lower boundary of the bin. When this bit is 1, the higher bin is output from the QA port and when it is 0, the lower bin is output. (Note that QB port always output contents of the addressed bin).

Some operation examples of the register-like storage block are presented in Fig. 5. In the table, the second column lists the output words from the FPGA module showing the test results. Each line represents a 250 MHz clock cycle. In each clock cycle, the functional block may perform NOOP, Book, Read or Refresh operation. In the event shown, we have written data A3, B3 into bin number 03, A4, B4 into 04, A5, B5, C5 into 05 and so on during the Book processes. Note that in this example, multiple data words may be written into a memory bin and the two words to be written into the same bin can be as close as in two adjacent clock cycles. With appropriately designed data forwarding unit, the functional block will update the addressed memory bin correctly regardless the number of clock cycles between the book operations.

The Read commend can be issued immediately after the last Book command. When a memory bin is read, the contents read out consist both the data written and the event ID (EVID) when the bin was updated. The stored EVID is compared with the current event ID and only if they are identical, as shown in line 21 to 27, the data bits stored in the bin is considered valid. If the stored EVID is different than the current event ID, the data in the bin will be treated as invalid and the bin is considered empty. For example, at line 31, 34 and 35, some non-zero numbers are read from corresponding bins, which are ignored since they are left over from old events.

| | Op | InData | Addr | Valid1 | EvtID | MemDataB 1 | 2 | 3 | MemDataA 1 | 2 | 3 |
|---|---|---|---|---|---|---|---|---|---|---|---|
| 0 | 00000FF300000000FF000000000000 | 0: NoOpRd | | 00 | 0 | 300 | | | | | | |
| 1 | 00000FF300000000FF000000000000 | 0: NoOpRd | | 00 | 0 | 300 | | | | | | |
| 2 | 00000FF300000000FF000000000000 | 0: NoOpRd | | 00 | 0 | 300 | | | | | | |
| 3 | 00000FF300000000FF000000000000 | 0: NoOpRd | | 00 | 0 | 300 | | | | | | |
| 4 | 00000FF300000000FF000000000000 | 0: NoOpRd | | 00 | 0 | 300 | | | | | | |
| 5 | 30000FFB95000000FF000000000000 | 3: Refresh | | 00 | 1 | 395 | | | | | | |
| 6 | 2A505FFB960000A5FF0000A5000000 | 2: Book | A5 | 05 | 1 | 396 | | A5 | | | | |
| 7 | 2B505FFB9600A5B5FF00A5B5000000 | 2: Book | B5 | 05 | 1 | 396 | A5 | B5 | | | | |
| 8 | 2C505FFB96A5B5C5FFA5B5C5000000 | 2: Book | C5 | 05 | 1 | 396 | A5 | B5 | C5 | | | |
| 9 | 2A303FFB960000A3FF0000A3000000 | 2: Book | A3 | 03 | 1 | 396 | | A3 | | | | |
| 10 | 2A404FFB960000A4FF0000A4000000 | 2: Book | A4 | 04 | 1 | 396 | | A4 | | | | |
| 11 | 2B303FFB9600A3B3FF00A3B3000000 | 2: Book | B3 | 03 | 1 | 396 | A3 | B3 | | | | |
| 12 | 2A808FFB960000A8FF0000A8000000 | 2: Book | A8 | 08 | 1 | 396 | | A8 | | | | |
| 13 | 2B404FFB9600A4B4FF00A4B4000000 | 2: Book | B4 | 04 | 1 | 396 | A4 | B4 | | | | |
| 14 | 2A707FFB960000A7FF0000A7000000 | 2: Book | A7 | 07 | 1 | 396 | | A7 | | | | |
| 15 | 2A606FFB960000A6FF0000A6000000 | 2: Book | A6 | 06 | 1 | 396 | | A6 | | | | |
| 16 | 2B808FFB9600A8B8FF00A8B8000000 | 2: Book | B8 | 08 | 1 | 396 | A8 | B8 | | | | |
| 17 | 2A909FFB960000A9FF0000A900A3B3 | 2: Book | A9 | 09 | 1 | 396 | | A9 | | | A3 | B3 |
| 18 | 10000FF300000000FF000000000000 | 1: Read | | 00 | 0 | 300 | | | | | | |
| 19 | 10001FF301000000FF000000000000 | 1: Read | | 01 | 0 | 301 | | | | | | |
| 20 | 10002FF302000000FF000000000000 | 1: Read | | 02 | 0 | 302 | | | | | | |
| 21 | 10003FFB9600A3B3FF00A3B3000000 | 1: Read | | 03 | 1 | 396 | A3 | B3 | | | | |
| 22 | 10004FFB9600A4B4FF00A4B400A3B3 | 1: Read | | 04 | 1 | 396 | A4 | B4 | | | A3 | B3 |
| 23 | 10005FFB96A5B5C5FFA5B5C500A4B4 | 1: Read | | 05 | 1 | 396 | A5 | B5 | C5 | | A4 | B4 |
| 24 | 10006FFB960000A6FF0000A6A5B5C5 | 1: Read | | 06 | 1 | 396 | | A6 | | A5 | B5 | C5 |
| 25 | 10007FFB960000A7FF0000A70000A6 | 1: Read | | 07 | 1 | 396 | | A7 | | | | A6 |
| 26 | 10008FFB9600A8B8FF00A8B80000A7 | 1: Read | | 08 | 1 | 396 | A8 | B8 | | | | A7 |
| 27 | 10009FFB960000A9FF0000A900A8B8 | 1: Read | | 09 | 1 | 396 | | A9 | | | A8 | B8 |
| 28 | 1000AFF30A000000FF0000000000A9 | 1: Read | | 0A | 0 | 30A | | | | | | A9 |
| 29 | 1000BFF30B000000FF000000000000 | 1: Read | | 0B | 0 | 30B | | | | | | |
| 30 | 1000CFF30C000000FF000000000000 | 1: Read | | 0C | 0 | 30C | | | | | | |
| 31 | 1000DFF3630000BFFF000000000000 | 1: Read | | 0D | 0 | 363 | | BF | | | | |
| 32 | 1000EFF30E000000FF000000000000 | 1: Read | | 0E | 0 | 30E | | | | | | |
| 33 | 1000FFF30F000000FF000000000000 | 1: Read | | 0F | 0 | 30F | | | | | | |
| 34 | 10010FF36A00006FFF000000000000 | 1: Read | | 10 | 0 | 36A | | 6F | | | | |
| 35 | 10011FF367000068FF000000000000 | 1: Read | | 11 | 0 | 367 | | 68 | | | | |
| 36 | 10012FF312000000FF000000000000 | 1: Read | | 12 | 0 | 312 | | | | | | |
| 37 | 10013FF313000000FF000000000000 | 1: Read | | 13 | 0 | 313 | | | | | | |
| 38 | 30000FFB96000000FF000000000000 | 3: Refresh | | 00 | 1 | 396 | | | | | | |
| 39 | 2D121FFB970000D1FF0000D1000000 | 2: Book | D1 | 21 | 1 | 397 | | D1 | | | | |
| 40 | 2D222FFB970000D2FF0000D2000000 | 2: Book | D2 | 22 | 1 | 397 | | D2 | | | | |

Fig. 5. Some examples of the operations of the register-like storage block

The last three column contains output data from memory port A for boundary coverage. In this example, the memory bin at BIN_NB -1 is addressed in each read operation. For example, at line 22, a read bin 04 command is issued, stored data in both bin 04 and 03 are output to port B and port A, respectively.

Once all Read commands are issued for an event, a Refresh command can be issued immediately as shown in this example. The Refresh command uses only one clock cycle to overwrite a rotationally chosen memory location and to increase the event ID counter. At line 38, the Refresh command increased EVID from 396 to 397. A new Book operation of the next event can be started without wasting any clock cycles.

As a further verification of the design, several thousands of events are processed through the sequence of Refresh, Book and Read operations. Along with the current event, several recent events and several very old events are selected and plotted in Fig. 6.

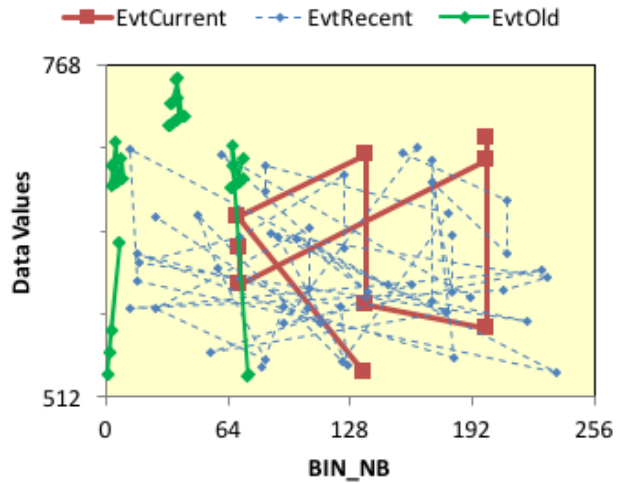

Fig. 6. Visualization of several events booked into the memory

For convenience of visualization, the bin number and data values written into the memory bins are used as X and Y coordinates in the plot. Each dot represents a stored data and the data points belonging to an event are connected.

The stored data has a range of 0 to 255 but a common constant is added to all data point while making the plot. (For the plot above the constant is 512.)

After finish booking all data points in the current event, data in all 256 memory bins are read out and plotted as shown in Fig. 7.

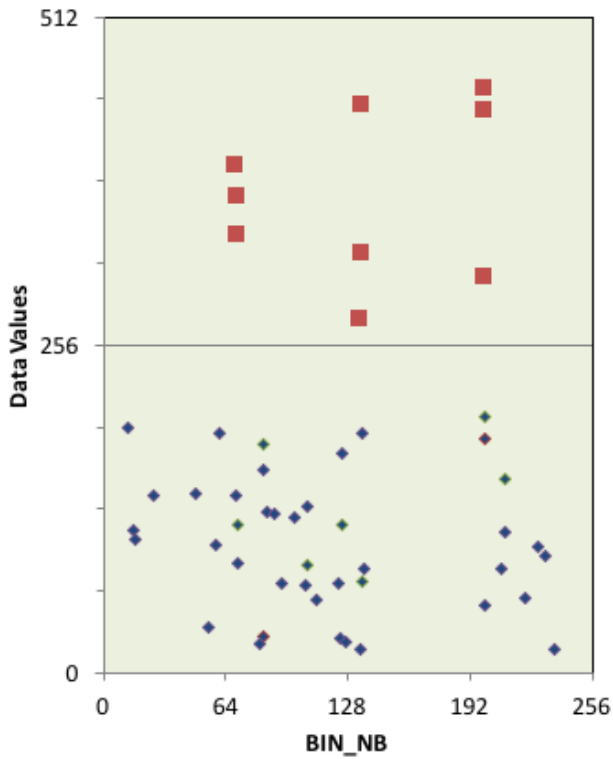

Fig. 7. Data points stored in the memory bins

It can be seen that all data points written during the current event and most points for recent event are seen in the memory bins. The memory bins containing data from current event can be easily identified. These data points identified as belonging the current event are added by a constant 256 in the spread sheet while making the plot to show them in the top portion of the plot for clarity.

Also, note that all data points written during the old events do not exist since they are erased in the refresh processes. Some points from the recent events may also be erased.

To study the refreshing process, a few thousand events are processed in the register-like storage block and the lower 9 bits of event ID in all 256 bins are plotted as shown in Fig. 6.

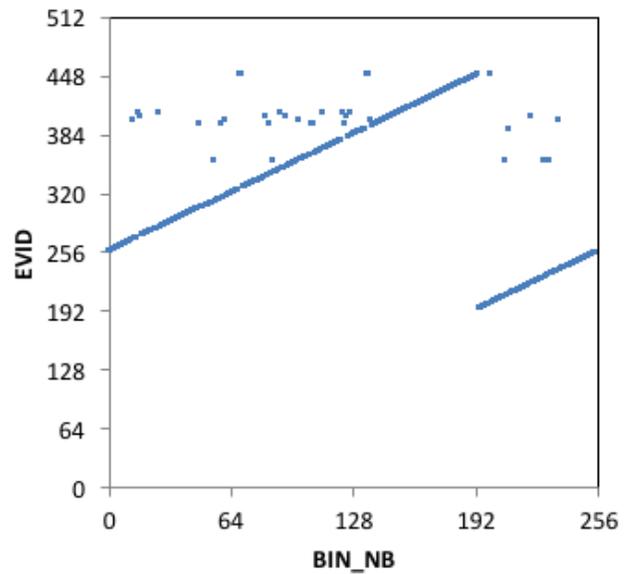

Fig. 6. Event ID stored in 256 bins

In this example, current event ID is 450 and highest dots in the plot represent data being written in the current event.

In every event, the refresh operation overwrites 1 bin rotationally selected from 256 bins. Therefore, the oldest event ID that exists in the entire memory is at most 256 events earlier than the current event. Some bins may contain newer event ID if they are used in the past 256 events. If we store 9 or more bits as event ID, the counter rollover will not cause mistakes during the event ID comparison. Similarly, at least 10, 11 or 12 bits should be used if the block RAM has 512, 1024 or 2048 bins.

IV. NOTES ON SEVERAL PRACTICAL APPLICATIONS

The register-like storage block is a building block that can be utilized for various functions in high energy physics trigger or offline event selection systems. In this section, we discuss several examples of practical applications.

A. The Cluster buffers

The firmware described in previous sections is essentially a cluster buffer. Data in an event are first written into bins in the cluster buffer and read out later. Depending on distribution of the hits, more than one hit may be stored in a bin, but not all bins are written or read in an event.

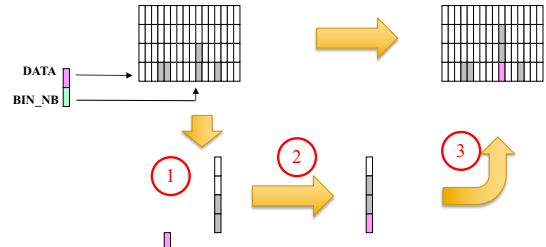

Fig. 7. The data update operation for the cluster buffer

One possible design of the data update unit in the cluster buffer is a data pusher as shown in Fig. 7, which saves current data into the lower bits while pushing the old data to higher bits.

In our design, we reserved memory space allowing up to three hits to be saved in a bin, but obviously the users may assign different capacities in their own applications.

An issue that the designers must consider is the exception handling. In other words, in some unlikely cases, if too many hits are to be filled into a bin causing it to full, how the overflow should be handled. A simplest approach is to do nothing, and the newest data will be kept while the oldest data are pushed out of the storage. It is also possible to implement additional logic so that oldest data are kept while newest data are dropped. If it is necessary, the users are allowed to design more sophisticated exception handling logic.

The contents being stored in the cluster buffer can be actual hit data or in most cases, index to the hit data buffer.

*B. Histograms*

A histogram can be viewed as a set of counters, one counter per bin, which increase by 1 when the bin is index during booking phase. The data updating operation of an online histogram unit is shown in Fig. 8.

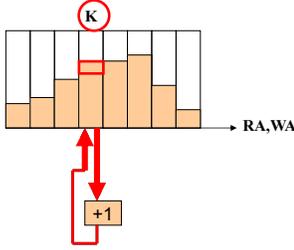

Fig. 8. The data update operation for the online histogram unit.

Of course, one may use logic elements in FPGA to implement an array of counter as online histogram but that will consume large amount of silicon resource. The register-like storage block can be turned into an online histogram unit with a minor modification of the data update unit. Simply put an adder in the data update unit and add the input number by 1 will fulfill the logic function for histogram.

In most time, an exception of overflow in each bin must be handled appropriately in the histograms. Usually, let the adder stop adding the input number by 1 when the input is bigger than certain value would be sufficient.

Implementing multiple dimension histogram is straightforward. For example, a 2048 bin 1D histogram can be used as a 32x64 2D histogram by simply assigning lower 5 address lines to the first dimension and higher 6 address lines to the second dimension.

In some situations, the online histogram may have less strict demands than the register-like storage block we discussed. For example, the incoming data may not come every clock cycle but rather, it may take at least a few clock cycles for the front-end circuit to produce one. In some applications the histograms may not need to be reset within a single clock cycle but rather, the users have enough time to readout the entire content of the memory and write zero into all locations. In this case, the histogram circuit can be simplified by eliminating some functional block in the register-like storage block.

*C. The Hough Transform Accumulator Array*

The Hough transform is a useful tool for track segment seeding in multi-layer detectors. In the Hough transform, a point (a hit on a detector layer) in physical coordinate space is transformed into a curve (or straight line in many cases) in track parameter space. Hits on different detector layers generated by the same track map to different curves on the parameter space, and the curves intersect to a spot representing the same set of track parameters.

In each detector/track configuration, usually there are several possible choices of Hough transform parameters and to certain extent they are mathematically equivalent. However, an appropriate set of parameters will help optimize performance in practical implementations and an example is shown in Fig. 9.

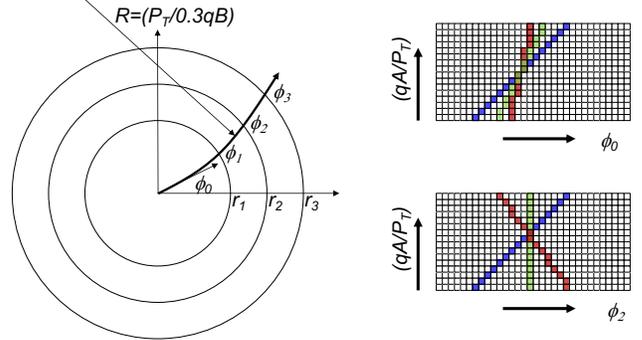

Fig. 9. An example of parameter selections in Hough transform

In a typical barrel detector in a solenoidal magnetic field, track curvature or inverse transverse momentum ($qA/P_T$) and track initial angle ($\phi_0$) are usually chosen as the Hough transform parameters. The hits in different detector layers are mapped into straight lines in the Hough transform space as shown in top-right diagram in Fig. 9. The inclined angles of these straight lines in actual implementation are determined by the bin width of the parameters and they confined within 45 to 90 degrees. In this situation, the intersections of these straight lines are blurred that causes track segment seed performances less optimal.

If we replace the parameter ($\phi_0$) with the coordinate of hit on layer 2 ($\phi_2$), the intersection in the Hough transform space will be a lot sharper as shown in lower-right diagram in Fig. 9. This selection will help reduce fake track segment rate when an event contains large number of hits. Similarly, the parameter choice of ($\phi_{58}$) given in Reference [11-12] helps improve the segment seeding performance.

In hardware implementation, an array of accumulators is needed and the accumulators in the cells along the curve will increase by 1 for an incoming hit. Ironically, the accumulators can be implemented with FPGA logic elements, but it would consume large silicon area and increase the system cost. The register-like storage block is a suitable alternative scheme featuring significant lower resource usage. We will discuss an example taken from our recent work as shown in Fig. 10.

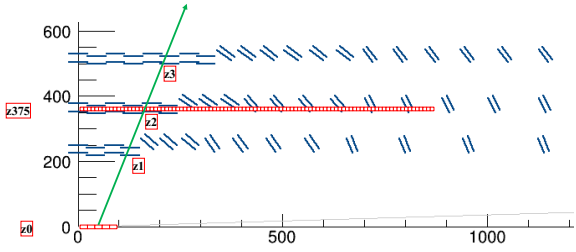

Fig. 10. The detector model and Hough transform parameters

The Hough transform block is a component in an exercise of implementation of a 3D track segment seeding engine. The detector model has the same diameters and lengths as the CMS Outer tracker except the tilting of the silicon sensor modules. The Hough transform block processes hits in r-z view with parameters z0 which is the z coordinate of the track starting point along the beam line and z375 which is approximately the z coordinate of the layer 2 hit. The choice of parameter z375 ensures that a sharp intersection of straight lines corresponding to layer 1, 2 and 3 hits in the Hough transform space, as discussed above. Note that the straight line corresponding to a layer 2 hit is a 90-degree vertical line in the Hough transform space.

The actual firmware implementation base on the register-like storage block is shown in Fig. 11.

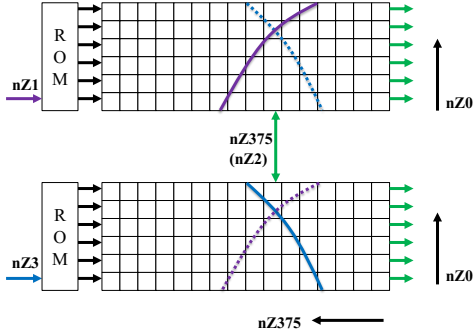

Fig. 11. The block diagram of the Hough transform storage

Two sets of Hough transform space are implemented each handling hits from layer 1 and layer 3 of the detector. Each horizontal row in the diagram represents a register-like storage block which contains 256 bins to cover the z375 dimension. Each register-like storage block represents a z0 bin and in our case, we have 10 z0 bins total.

For each hit in layer 1 or 3 with integer coordinates nZ1 or nZ3, all RAM blocks for different nZ0 are written at different addresses (nZ375), which corresponding to a straight line in the parameter plane. The addresses are pre-calculated reflecting the slopes of the straight lines and are stored in a set of ROM lookup tables. The scheme of using lookup tables will accommodate any shapes of the curves beyond straight lines to maintain flexibility in case the Hough transform engine is to be used for more complicate detector geometries as in the examples described in References [6-10].

The contents of the cells in the Hough transform space are bit patterns as shown in Fig. 12.

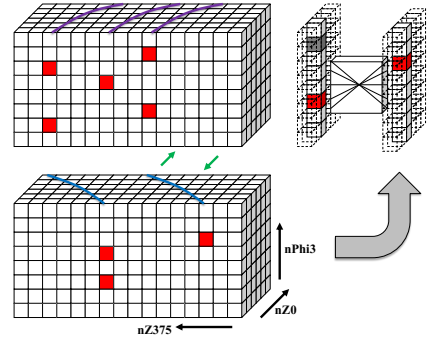

Fig. 12. The contents of the Hough transform cells and the Tiny Triplet Finder

The payload in each cell is a 128-bit word which represents a bit pattern spanning in $\phi$ dimension. For any hit in layer 1 or layer 3, the $z$ coordinate is used to address the RAM location and the $\phi$ coordinate is used to set a particular bit in the 128-bit words for all the RAM blocks. Note that this bit setting is an update operation meaning that while the selected bit is set, the other bits remain unchanged.

During the readout phase, hits in layer 2 are cycled through the Hough transform block. The $z$ coordinate of the layer 2 hits is converted into addresses (nZ2) of the RAM blocks. The conversion can be done in the ROM lookup tables also for maximum flexibility, but in our application, nZ2 is nearly the same as nZ375 and therefore the RAM blocks are addresses directly by nZ2. As pointed out earlier, a layer 2 hit represent a vertical line in the Hough transform space.

For each layer 2 hit, two sets of bit patterns are output from the two Hough transform blocks. The bit patterns with the same nZ0 (plus boundary overlap) from the two Hough transform blocks are brough together to feed the Tiny Triplet Finders. The Tiny Triplet Finders will search for the coincidence in the $\phi$ dimension which will be discussed in other papers. Note that the Hough transform is not used to complete the full segment seeding process, but rather, it is merely are feeder to another stage (Tiny Triplet Finder) so that constraints of track segments in 3 dimensions can be utilized for better fake segment rejection ratio.

## V. DISCUSSIONS

A register-like storage block is design, implemented and tested. Single clock writing, reading and refreshing performance allows wide applications in various high energy physics trigger systems.

The firmware scheme allows designers to insert more pipeline stages into complex combination logics. In our design, both the EVID Check Unit and the Data Update Unit are pipeline stages, respectively. This way, the operating speed degrading caused by the complex combination logics is negligible. The firmware operates at 250 MHz which is nearly the highest operating speed of the block RAM resource in this device.

The cluster buffers, histograms and Hough transform accumulators are several examples of the applications of the register-like storage blocks. Sometimes, the Hough transform accumulators are referred as "2D histogram" but there is a crucial difference between them. A histogram usually updates a single bin for each incoming hit while in Hough transform, a set of bins representing a straight line (or a curve in general) must be updated. This requirement forces us to use a set of register-like storage blocks to implement a Hough transform space. In our case described earlier, for example, we need 10 register-like storage blocks to implement a Hough transform space, for 10 z0 bins. Therefore, it will be very helpful to keep implementation details in mind during the algorithm design stage. Choosing less (coarser) bins in one dimension in the transform parameter space while using more (finer) bins in another dimension will reduce resource consumption in final FPGA firmware.

REFERENCES

[1] J. Wu, "Register-like Block RAM: Implementation, Testing in FPGA and Applications for High Energy Physics Trigger Systems," in *2016 IEEE Real Time Conference* 2016, available via: {https://indico.cern.ch/event/390748/contributions/1825169/}
[2] J. Wu, "Register-Like Block RAM with Boundary Coverage and Its Applications for High Energy Physics Trigger Systems," in *2018 IEEE Real Time Conference* 2018, available via: {https://indico.cern.ch/event/543031/contributions/2921544/attachments/1663394/2665714/RegisterLikeRAM2018b.pdf}
[3] H. Sadrozinski & J. Wu, "Applications of Field-Programmable Gate Arrays in Scientific Research", Taylor & Francis, December 2010.
[4] Mikael Martensson, "Fast pattern recognition with the ATLAS L1Track trigger for the HL-LHC," in *the 25th International workshop on vertex detectors* 2016, available via: {https://pos.sissa.it/287/069/}
[5] D. Cieri, *et al*., "Hardware demonstrator of a compact first-level muon track trigger for future hadron collider experiments," in *Journal of Instrumentation,* 2019, P02027, vol. 14, available via: {https://doi.org/10.1088/1748-0221/14/02/P02027}
[6] S. A. Rodenko, *et al*., "Track reconstruction of antiprotons and antideuterons in the coordinate-sensitive calorimeter of PAMELA spectrometer using the Hough transform," in *Journal of Physics: Conference Series,* 2019, vol. 1189.
[7] Z. Deng, *et al*., "Tracking within Hadronic Showers in the CALICE SDHCAL prototype using a Hough Transform Technique," in *Journal of Instrumentation,* 2017, P05009, vol. 12, available via: {https://iopscience.iop.org/article/10.1088/1748-0221/12/05/P05009}
[8] D Primor, *et al*., "A novel approach to track finding in a drift tube chamber," in *Journal of Instrumentation,* 2007, P01009, vol. 2, available via: {https://iopscience.iop.org/article/10.1088/1748-0221/2/01/P01009}
[9] W. Andersson, *et al*., "Reconstructing Hyperons with the PANDA Detector at FAIR," in *Journal of Physics: Conference Series,* 2016, vol. 742. available via: {https://iopscience.iop.org/article/10.1088/1742-6596/742/1/012009}
[10] L. Niu, *et al*., "Track reconstruction based on Hough-transform for nTPC," in *Chinese Physics C,* 2014, vol. 38. available via: {https://iopscience.iop.org/article/10.1088/1674-1137/38/12/126201}
[11] R. Aggleton, *et al*., "An FPGA based track finder for the L1 trigger of the CMS experiment at the High Luminosity LHC," in *Journal of Instrumentation,* 2017, P12019, vol. 12, available via: {https://doi.org/10.1088/1748-0221/12/12/P12019}
[12] C. Amstutz *et al*., "An FPGA-based track finder for the L1 trigger of the CMS experiment at the high luminosity LHC," *2016 IEEE-NPSS Real Time Conference (RT)*, 2016, pp. 1-9, doi: 10.1109/RTC.2016.7543102.